\documentclass[useAMS,usenatbib]{mn2e}
\def\aap{AA}

\def\apjl{ApJL}

\def\mnras{MNRAS}
\def\apj{ApJ}
\def\apjs{ApJS}

\def\aj{AJ}
\def\jcap{JCAP}

\def\ase{{\prime\prime}}

\usepackage{graphicx}
\usepackage{float}
\usepackage{amssymb}
\usepackage{amsfonts}
\usepackage{amsmath} 
\usepackage{color}
\def\aaemail{\tt aagnello@eso.org}
\def\eso{European Southern Observatory, Karl-Schwarzschild-Strasse 2, 85748 Garching bei M{\"u}nchen, DE}
\def\dark{Dark Cosmology Centre, Niels Bohr Institute, University of Copenhagen, Juliane Maries Vej 30, DK-2100 Copenhagen, Denmark}
\def\mi{Dipartimento di Fisica, Universit\`a  degli Studi di Milano, via Celoria 16, I-20133 Milano, Italy}
\def\ucla{Department of Physics and Astronomy, PAB, 430 Portola Plaza, Box 951547, Los Angeles, CA 90095-1547, USA}
\def\ucd{Department of Physics, University of California Davis, 1 Shields Avenue, Davis, CA 95616, USA}
\def\mpa{Max-Planck-Institut f{\"u}r Astrophysik, Karl-Schwarzschild-Str. 1, D-85741 Garching, Germany}
\def\iofa{Institute of Astronomy, University of Hawaii, 2680 Woodlawn Drive, Honolulu, HI 96822, USA}

\title[The new quasar quad J1433+6007]{Discovery and first models of the quadruply lensed quasar SDSS J1433+6007} 
\author[Agnello et al.]{
  Adriano Agnello$^{1}$\thanks{\aaemail},
 Claudio Grillo$^{2,3}$,
 Tucker Jones$^{4,5,\dag}$,
 Tommaso Treu$^{6,\ddag}$,\and
 Mario Bonamigo$^{3}$,  
 Sherry H. Suyu$^{7,8,9}$
  \medskip\\
  $^1$\eso\\
  $^2$\mi\\
  $^3$\dark\\
  $^4$\ucd\\
  $^5$\iofa\\
  $^6$\ucla\\
  $^7$\mpa\\
  $^{8}$Institute of Astronomy and Astrophysics, Academia Sinica, P.O.~Box 23-141, Taipei 10617, Taiwan\\
$^{9}$Physik-Department, Technische Universit\"at M\"unchen, James-Franck-Stra\ss{}e~1, 85748 Garching, Germany\\
  $^\dag$ Hubble Fellow, $^\ddag$ Packard Fellow.\\
}
\begin{document}

\voffset-.6in

\date{Accepted . Received }

\pagerange{\pageref{firstpage}--\pageref{lastpage}} 

\maketitle

\label{firstpage}

\begin{abstract}
We report the discovery of the quadruply lensed quasar J1433+6007, mined in the SDSS DR12 photometric catalogues using a novel outlier-selection technique, without prior spectroscopic or UV excess information. Discovery data obtained at the Nordic Optical telescope (NOT, La Palma) show nearly identical quasar spectra at $z_s=2.74$ and four quasar images in a fold configuration, one of which sits on a blue arc. The deflector redshift is $z_{l}=0.407,$ from Keck-ESI spectra. We describe the selection procedure, discovery and follow-up, image positions and $BVRi$ magnitudes, and first results and forecasts from simple lens models.

\end{abstract}

\section{Introduction}

Strong gravitational lensing by galaxies enables the study of distant sources and luminous and dark matter in galaxies over a range of redshifts. When the source is a quasar, its multiple images give a wealth of information on: the source central engine and the stellar content of the lens, via microlensing by individual stars \citep{sch16,bra16}; substructure in the lens, via astrometric and flux-ratio `anomalies' \citep{dal02,nie14,agn17}; 
quasars and their hosts at $z_{s}\approx2$ \citep{rus14,agn16,din17}; and cosmological distances, from the time-delays between the light-curves of different images \citep{ref64,par09,suy14}.

However, strong lenses are rare, as they require the precise alignment of a distant source with (at least) a galaxy. Since quasars are rare objects themselves and are less frequent at higher redshift, quasar lenses are even rarer.  \citet{om10} estimate that $\approx0.2$ lensed quasars per square degree, brighter than $i=21,$ should be present in wide-field surveys, with a majority of doubly imaged quasars and $\approx20\%$ quadruples. Their predicted source redshifts are distributed at $z\approx3.0\pm0.3,$ higher than those of most quasar lenses discovered in wide-field searches, specifically in the SDSS Quasar Lens Search \citep[SQLS,][]{ogu06}, which targeted objects with quasar fibre-spectra, themselves based on UV excess (UVx) preselection.
Extensions of lens searches to higher completeness must rely solely on photometric preselection with limited UVx information. In view of current and upcoming wide-field surveys diverse techniques have been developed to this aim \citep[e.g.][]{agn15,sch16,mor16,ost17,wil17,lin17}.

\begin{figure*}
 \centering
 \includegraphics[width=0.95\textwidth]{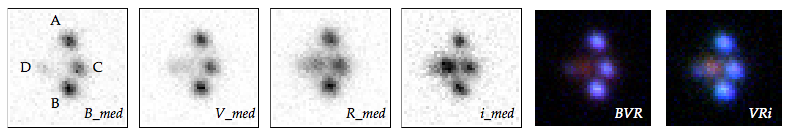}\\
 \includegraphics[width=0.3\textwidth]{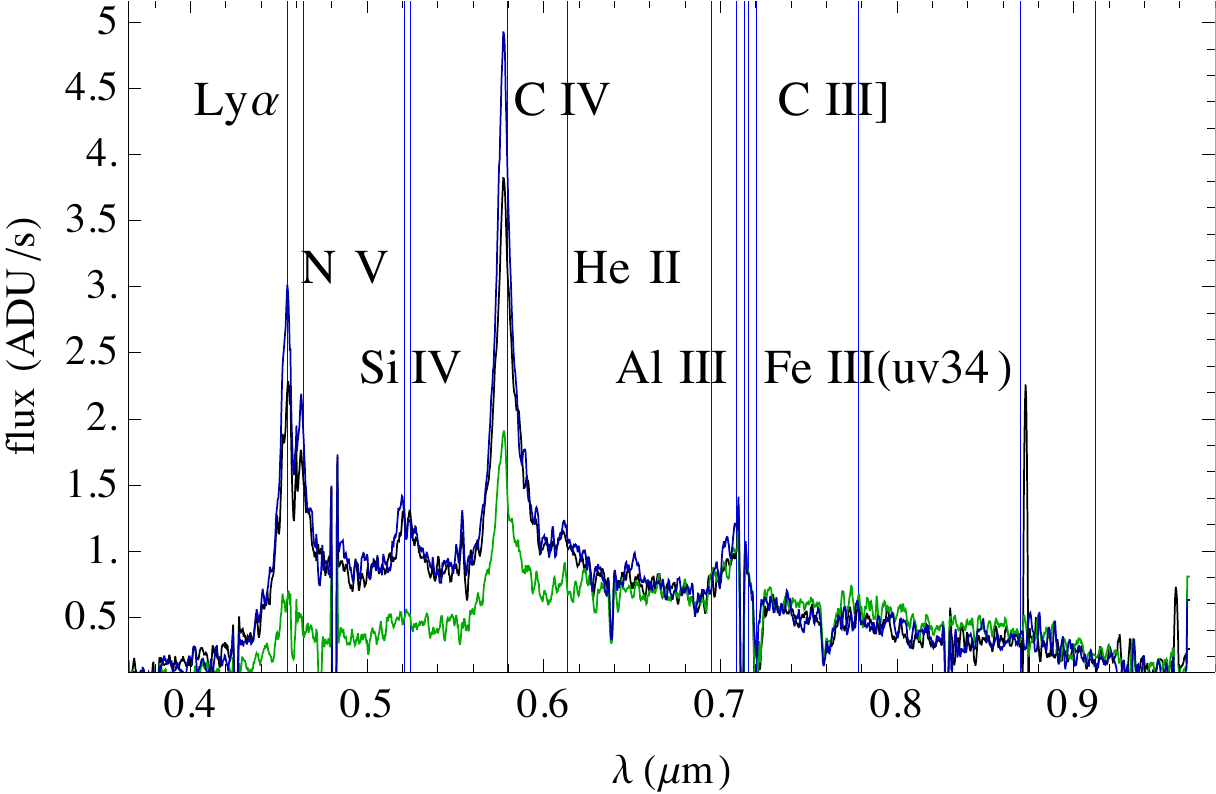}
 \includegraphics[width=0.3\textwidth]{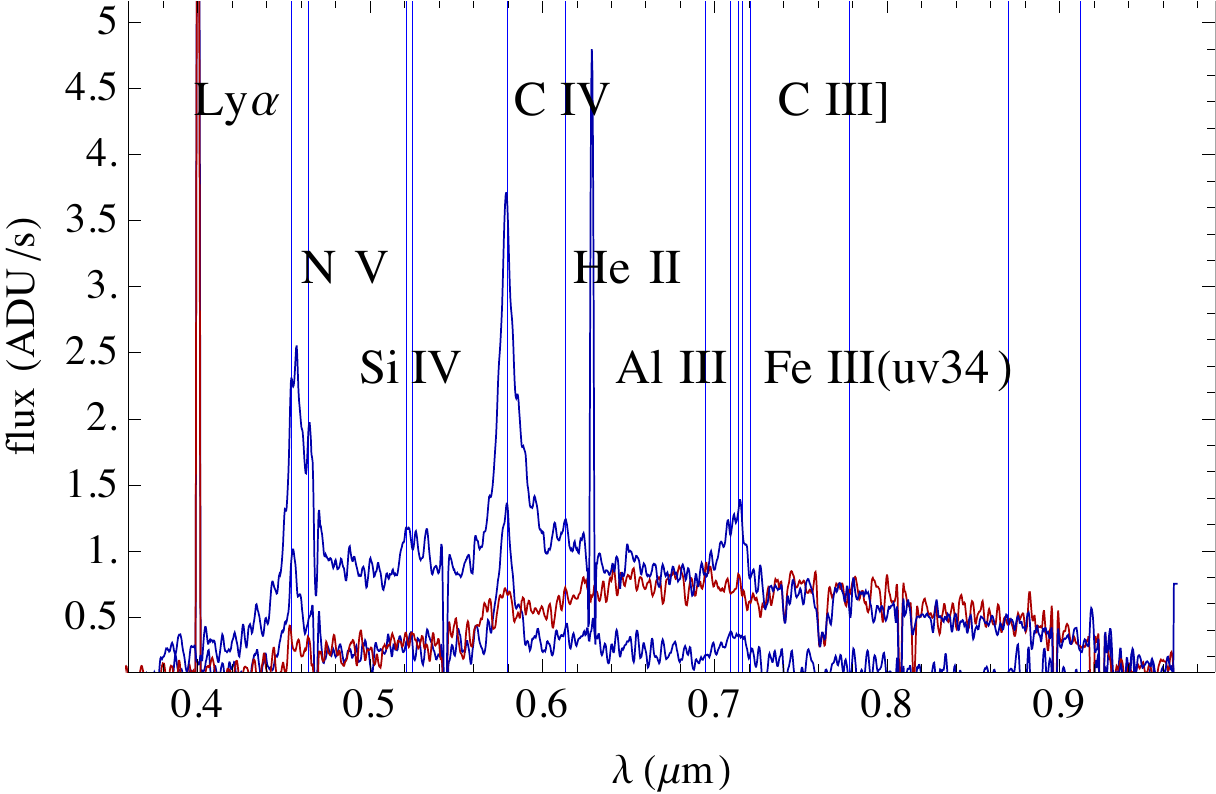}
 \includegraphics[width=0.3\textwidth]{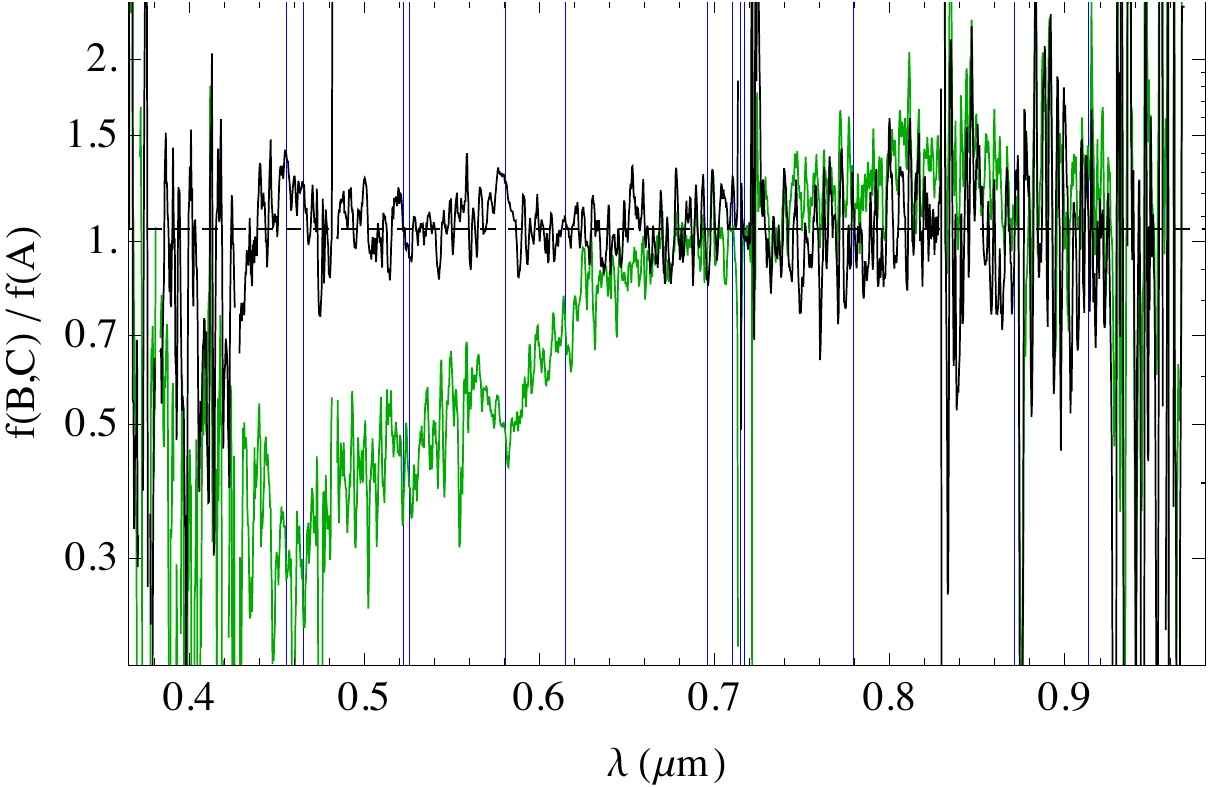}
 
\caption{{J1433+6007 discovery and follow-up data.
\textit{Top:} ALFOSC $BVRi$ median coadds and colour-composites, $0.21^{\ase}$/px, North up and East left. A faint arc, barely visible in $B-$band, emanates from the northern-most image. \textit{Bottom:} North-South (left) and East-West spectra (center), and flux-ratios of spectra in North-South slit (right). Image C leaks in the North-South slit and is significantly reddened (right, green line), whereas images A,B have almost indistinguishable spectra (right, black line). The source has $z_{s}=2.74,$ as evaluated from C~\textsc{iv}, where the wavelength calibration is most accurate.
 }}
\label{fig:puppies}
\end{figure*}

Here, we report on the discovery and follow-up of a quadruply lensed quasar, J1433+6007, mined with a novel technique in the Sloan Digital Sky Survey \citep[SDSS,][]{aba09} DR12 footprint without spectroscopic or UVx information. 
The two brightest images are separated by $\approx3.6^{\ase},$ the source redshift from the discovery spectra is $z_{s}=2.74,$ and the lens redshift is $z_{l}=0.407$ from follow-up Keck-ESI spectroscopy.
We describe the target/candidate selection procedure in Section \ref{sect:cands}, the confirmation and follow-up of J1433 in Sect.~\ref{sect:fu}, and first lens models in Sect.~\ref{sect:mods}. We conclude in Section~\ref{sect:conc}. In what follows, SDSS magnitudes are in the AB system, WISE \citep{wri10} magnitudes in the Vega system, and where necessary we adopt concordance cosmological parameters $\Omega_{\Lambda}=0.7,$ $\Omega_{m}=0.3,$ $H_{0}=70$km/s/Mpc.

\section{Candidate Selection}
\label{sect:cands}

Since half of the known quasar lenses in SDSS have extended morphology, due to the presence of the deflector  \citep[see][for a discussion]{wil17}, objects with $\log_{10}\mathcal{L}_{star,i}<-11$ or \texttt{psf{\_}i}-\texttt{mod{\_}i}$>$0.075, \texttt{mod{\_}i}$<20.5,$ and WISE $W1-W2>0.55,$ $W2-W3<3.1+1.5(W1-W2-1.075)$ were pre-selected. Minimal $griz$ cuts were used. The $i-$band selection is used as a morphological preselection, whereas the WISE cuts are an extension of those by \citet{ass13} to exclude most quasars at $z<0.35$ and narrow-line galaxies.
Quasar lens \textit{targets} were then selected based on their catalog magnitudes, then visually inspected to  exclude obvious contaminants, yielding the final \textit{candidate} sample.

Lensed quasars are rare among quasars, which in turn are rarer than blue galaxies, hence we used a novel outlier selection to mine targets (detailed elsewhere, Agnello 2017), retaining peculiar objects by excluding more common ones.
When tested on the SQLS morphologically-selected lens candidates of \citet{ina12}, this procedure recovered 9 of the 10 lenses and excluded half of the 40 false positives, without any UVx or fibre-spectroscopy information.

Four classes of `common' objects were defined, roughly corresponding to nearby ($z<0.75$) quasars, isolated quasars at higher ($z\approx2$) redshift, blue-cloud galaxies and faint ($W2\gtrsim15$) objects. Each class $k$ was represented by a single Gaussian with mean $\boldsymbol{\mu}_{k}$ and covariance $\mathbf{C}_{k}$ in a space given by $g-r,$ $g-i,$ $r-z,$ $i-W1,$ $W1-W2,$ $W2-W3,$ $W2$ and each object $\mathbf{f}$ was assigned pseudo-distances defined as $d_{k}=0.5\left\langle\mathbf{f}-\boldsymbol{\mu}_{k},\mathbf{C}_{k}^{-1}(\mathbf{f}-\boldsymbol{\mu}_{k})\right\rangle.$ Objects that were `far' enough from the four class centers, based on linear combinations of their $d_{k}$ values, were retained as targets.

This yielded $\approx250$ candidates brighter than $i=20.0$ over the whole SDSS-DR12 footprint, of which $\approx40$ known quasar lenses or pairs. J1433+6007, at r.a.= 14:33:22.8, dec.=+60:07:13.44 (J2000), showed two well-separated blue images on either sides of two red objects, blended in three photometric components by the SDSS pipeline.

\begin{table*} 
\centering
\begin{tabular}{lc|cc|cccc|ccl}
\hline
img.	&	$\delta x(^\ase)$	&	$\delta y(^\ase)$	&	$B$	&	$V$	&	$R$	&	$i$ & $\mu$ & $t-t_{A}$ \\
	&	$=-\cos(\mathrm{dec.})\delta\mathrm{r.a.}$	&	$=\delta\mathrm{dec.}$	&	(mag)	&	(mag)	&	(mag)	&	(mag) &  & (days)\\
\hline
A & $0.00\pm0.025$ & $0.00\pm0.025$ & $20.26\pm0.04$ & $19.78\pm0.01$ & $19.26\pm0.01$ & $19.32\pm0.01$ & $2.62$ & 0.00 \\ 
B & $-0.070\pm0.025$ & $-3.650\pm0.025$ & $20.09\pm0.03$ & $19.63\pm0.01$ & $19.13\pm0.01$ & $19.10\pm0.01$ & $3.57$ & 15.0 \\ 
C & $0.766\pm0.025$ & $-2.056\pm0.025$ & $20.50\pm0.05$ & $19.92\pm0.01$ & $19.30\pm0.01$ & $19.14\pm0.01$ & $-3.07$ & 25.0 \\ 
D & $-2.138\pm0.050$ & $-2.132\pm0.050$ & $22.00\pm0.14$ & $21.30\pm0.02$ & $20.63\pm0.02$ & $20.38\pm0.01$ & $-0.62$ & 113.0 \\ 
G & $-1.152\pm0.025$ & $-1.950\pm0.025$ & $21.87\pm0.15$ & $20.69\pm0.12$ & $19.39\pm0.01$ & $18.52\pm0.01$ & ---  & ---\\ 
\hline
\end{tabular}
\caption{Image positions, $BVRi$ magnitudes, and model-predicted values of magnifications and time-delays. The nominal errors on some magnitudes would be smaller than quoted, but are limited by the accuracy of the ALFOSC zero-points and observing sky conditions.}
\label{tab:bigtab}
\end{table*}

\section{Follow-up}
\label{sect:fu}
Long-slit discovery spectra were obtained on 2017/01/19,20 as part of a candidate lens follow-up program (P42-019, PI Grillo). We used the Andalucia Faint Object Spectrograph and Camera (ALFOSC) at the 2.5m Nordic Optical Telescope (NOT) in La Palma (Spain), and the $1^\ase-$wide long-slit with the {\#}4 grism, covering a wavelength range $3200\rm{\AA} <\lambda< 9600\rm{\AA}$ with a dispersion of $3.3 \rm{\AA}/\rm{pixel}$. Standard \textsc{Iraf} routines were used for bias subtraction, flat-field corrections and wavelength calibration. From ALFOSC $BVRi$ imaging, we obtained the positions and magnitudes subsequently used for lens models. Deeper, high-resolution spectroscopy was obtained with the Echellette Spectrograph and Imager (ESI) at the Keck \textsc{II} telescope on 2017/01/20 (PI Jones), with a $1^\ase$-wide slit, and reduced with ESIRedux\footnote{Available at \texttt{http://www2.keck.hawaii.edu/inst/esi/ESIRedux/}}.
 
In what follows, quasar images are labeled as shown in Fig~\ref{fig:puppies}, top-left, along the expected arrival times. ALFOSC pixels measure $0.21^{\ase}$ per side.

\subsection{NOT discovery and follow-up spectra}
We took two 600s exposures with the slit aligned North-South, through 14:33:22.8+60:07:13.44, and one (900s) with East-West alignment, through 14:33:22.8+60:07:14.5. This enabled simultaneous spectroscopy of the two prominent quasar images and the two red objects, respectively. Arc (HeNe, Ar) and flat lamps were used for calibrations.

The North-South spectra show three nearly identical traces corresponding to the same $z_{s}=2.74$ quasar (Fig.~\ref{fig:puppies}, lower-left panel). The bright, outer traces correspond to the blue images (A,B) visible in the SDSS. The central, fainter trace is given by a third quasar image C, corresponding to the West-most red object, just outside the slit, thus confirming J1433 as a multiply imaged quasar. The East-West spectra  (Fig.~\ref{fig:puppies}, lower-middle panel) show clear traces corresponding to images C,D and the lens galaxy (G). Images A,B have almost indistinguishable spectra, with uniform flux ratio $\approx1.05,$ whereas image C undergoes substantial extinction bluewards of C~\textsc{iii}$\left.\right]$. Micro-lensing results in $\approx5-10\%$ flux-ratio differences between the continua and emission lines  (Fig.~\ref{fig:puppies}, lower-right panel).

\subsection{NOT imaging}
Follow-up imaging data with good seeing ($\approx0.6^{\ase}$ FWHM at Zenith) were obtained with ALFOSC in $B,$~$V,$~$R,$~$i$ bands, using multiple $60$s exposures per band. 
 Standard \textsc{Python} routines were used for bias subtraction, flat-fielding and coadding. The median coadds and colour-composites are shown in Fig.~\ref{fig:puppies}, where Ly~$\alpha$ from the quasar images dominates in $B-$band and the lens brightens up in redder bands.
 
The relative displacements are obtained both from imaging and spectroscopic data. Individual traces in the spectra are well described by Gaussians in the spatial direction, whose run with wavelength can be modeled with uncertainties as low as 0.125px$=25$mas. Uncertainties from imaging-only data, though nominally smaller, are dominated by systematics from different noise realizations. Residuals between imaging data and model, mostly due to faint features and PSF mismatch, are within the noise level.

Table~1 gives the positions of the four images (A,B,C,D) and deflector (G), relative to image A, and $BVRi$ magnitudes. Images C and D are substantially reddened, and blending between D and G is significant in $B-$band.

\begin{figure}
 \centering
 \includegraphics[width=0.45\textwidth]{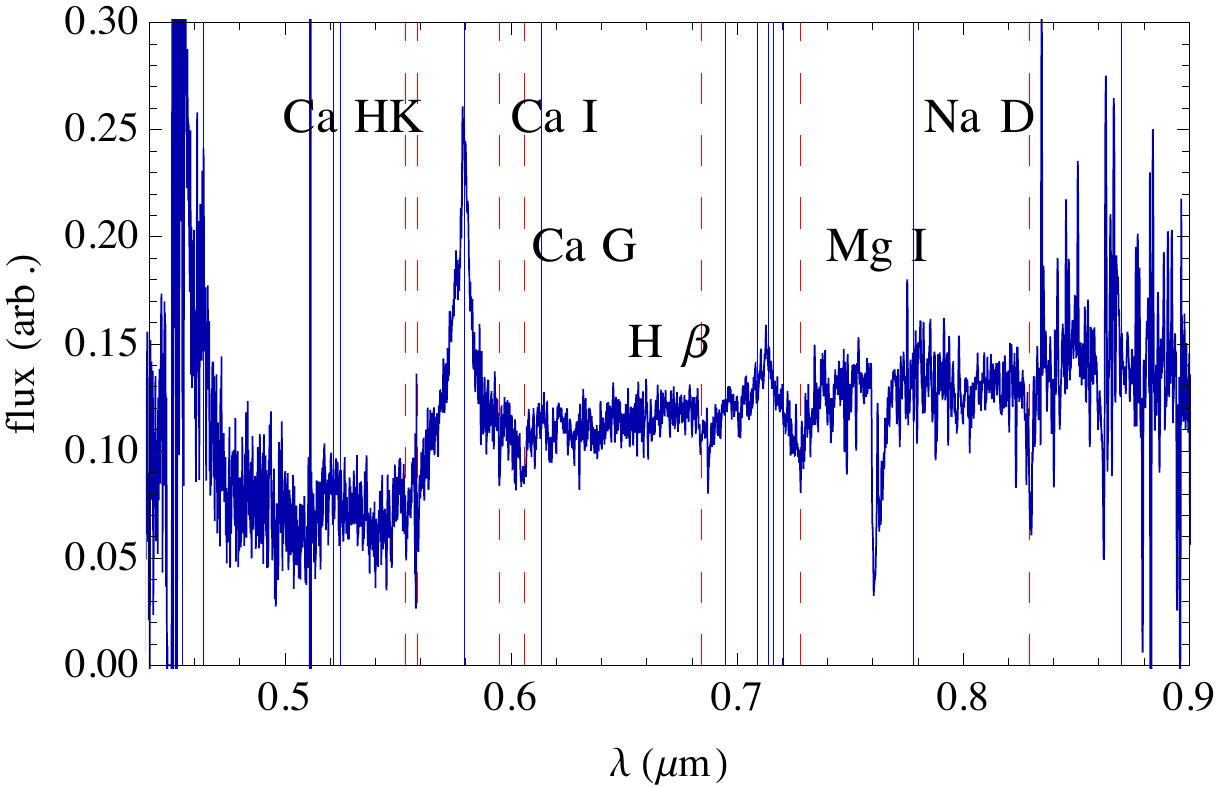}\\
\caption{{ESI follow-up spectra, with prominent absorption features at $z_{l}=0.407,$ which we associate with G. \textit{Red dashed} lines: Ca~HK, G-band $\lambda$4304, Mg~\textsc{i}, Na~D absorption at $z_l$. \textit{Blue solid} lines: quasar emission at $z_s$. The Ca~HK complex is visible, albeit at lower S/N, also in the ALFOSC spectra. }}
\label{fig:ESI}
\end{figure}

\subsection{Keck-ESI follow-up}
ESI spectra of G were taken in echellette mode, with $1.0^{\ase}$ slit-width oriented East-West. The total integration time was 45 minutes split into 3 exposures of 900 seconds each. The combined 1D spectrum of G and D, shown in Figure~2, has distinctive absorption features at $z_{l}=0.407$ (Ca~HK, G-band, H$\beta,$ Mg~\textsc{i}/Fe complex, Na~D), which we associate with the lens galaxy. The same features could be seen in the ALFOSC discovery spectra, though not as clear.

\begin{table} 
\centering
\begin{tabular}{lc|ccccc|}
\hline
  &  $\theta_{\rm E}$ & $q$  & $\phi_{l}$  & $\phi_{\rm s}$ & $\gamma_{\rm s}$ \\
\hline
best  &  1.80$^\ase$ & 0.50  & 0.18[rad]  & 1.12[rad] & 0.10  \\
68\% low  &  1.70$^\ase$ & 0.43  & 0.13[rad]  & 0.96[rad] & 0.08  \\
68\% high  &  1.90$^\ase$ & 0.58  & 0.25[rad]  & 1.22[rad] & 0.13  \\
\hline
\end{tabular}
\caption{Lens model parameters: best-fit (first column) and 68\% confidence intervals, marginalized over other parameters. Tight (and expected) degeneracies among parameters are present, given in eq.~(\ref{eq:degen}), except for the combination $\theta_{\rm E},$ which corresponds to the Einstein radius. Angles $\phi_{l},\phi_{\rm s}$ are positive counter-clockwise from West.}
\label{tab:lenstab}
\end{table}

\begin{figure}
 \centering
 \includegraphics[width=0.45\textwidth]{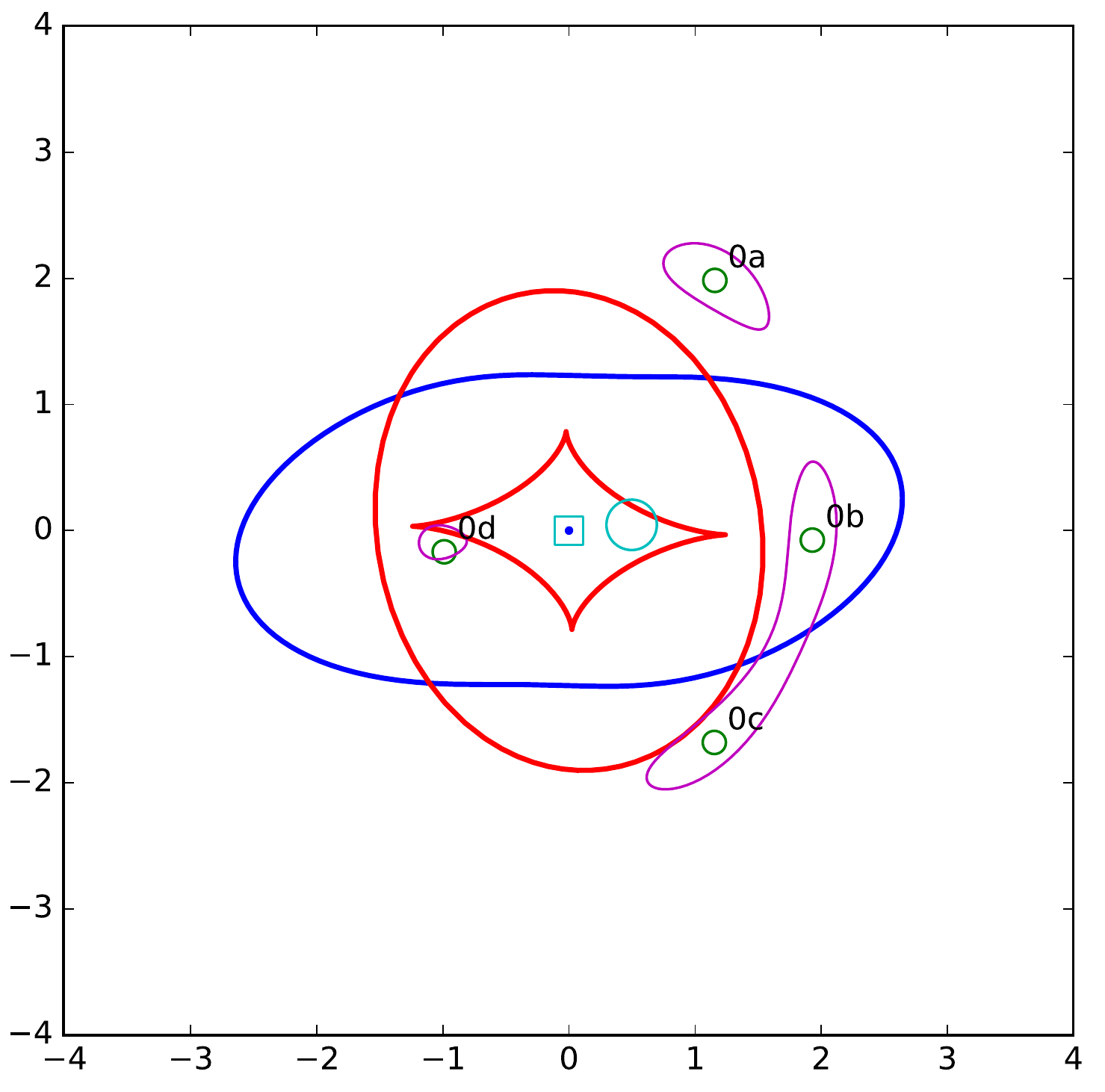}\\
\caption{{Lens model results: input images (green circles) and lens center (cyan square), $0.2^\ase-$radius circle around the inferred source position (cyan), critical curve (blue) and radial and astroid caustics (red). Purple lines show isophotes corresponding to the source-plane contour. The position of image D is slightly biased, due to blending with G, but is recovered within the adopted uncertainties. }}
\label{fig:GLEE}
\end{figure}

\section{Lens models}
\label{sect:mods}
From the relative displacements in Table~1, we fit a simple lens model to obtain the enclosed (2D) mass and predicted magnifications and time-delays. We do a conjugate-point analysis using \textsc{Glee} \citep{suy10,suy12}, adopting $25$mas uncertainties on the positions of A,B,C and G, and $50$mas uncertainties on the position of image D. The surface density of the lens is described by
%
\begin{equation}
\kappa(x,y)=\frac{\Sigma(x,y)}{\Sigma_{\rm cr}} =\frac{b}{2\sqrt{X^{2}/(1+e)^{2}\, +Y^{2}/(1-e)^{2}}}
\end{equation}
\citep{kas93}, along the principal axes $(X,Y)$ of G, where $\Sigma_{\rm cr}=\mathrm{c}^{2}D_{s}/(4\pi G\, D_{l}D_{ls})$ factors the dependence on angular-diameter distances.
 The Einstein radius is defined as the geometric mean of the major and minor axes of the critical curve, $\theta_{\rm E}=2b\sqrt{q}/(1+q).$ We include external shear, with amplitude $\gamma_{\rm s}$ and angle $\phi_{\rm s},$ and let all parameters free to vary, including the lens flattening $q=(1-e)/(1+e)$ and position angle $\phi_{l}.$

The results are summarized in Tables~1,2 and Figure~3. The Einstein radius is robustly determined to $\theta_{\rm E}=(1.80\pm0.10)^\ase$, very close to half the A-B separation and independent of other inferred quantities. Flux-ratios from the best-fit model are comparable to those measured in $i-$band, accounting for differential extinction whose existence in lensing is well established \citep[e.g.][]{fal99,med05,agn17}. The predicted delays $t_D-t_A>100$d, $t_C-t_A\approx25$d can be accurately measured in one or two seasons of high-cadence monitoring.

Monopole-quadrupole degeneracies \citep[e.g.][]{koc06} are present, in particular
\begin{eqnarray}
\nonumber \gamma_{\rm s} & \approx & 0.10+0.35(q-0.5),\\
\nonumber \phi_{\rm s} & \approx &  1.1-1.5(b-2.5),\\
 \gamma_{\rm s} & \approx &  0.10+0.35(\phi_{l}-0.8)
 \label{eq:degen}
\end{eqnarray}
for this system. These may be broken with: observations of the line-of-sight environment, to characterize external contributions to the deflections; and higher-resolution imaging, of both the lensed quasar host and of G, to disentangle shear and lens flattening. 
  
The velocity dispersion $\sigma$ of stars in the lens is a useful observable that can be estimated from the lens model itself. In fact, a direct measurement of $\sigma$ and its comparison with predictions from lensing can be used to measure cosmological distances and to constrain the dark matter density profile of the lens \citep{tk04,gri08,par09,suy14,son15,jee16}.  In the Singular Isothermal Sphere (SIS) limit $q\rightarrow1,$ $\sigma$ depends weakly on location\footnote{The details depend on the steepness of the starlight's profile and the orbital anisotropy \citep{bar11,agn13}.}
~and is well approximated by
\begin{equation}
\sigma_{\rm sis}^{2}=\frac{\mathrm{c}^{2}\theta_{\rm E}D_{s}}{4\pi D_{ls}}\ .
\label{eq:SIS}
\end{equation}
 With the above values, we then obtain $\sigma=(290\pm8)$ km/s, as expected for massive ellipticals \citep{tre05}.
 A direct comparison between the estimate in eq.~(\ref{eq:SIS}) and direct measurements will need dynamical models that encompass asphericity and inclination effects \citep{bar11}.

\section{Discussion and Conclusions}
\label{sect:conc}
We have found a new, quadruply lensed quasar at RA= 14:33:22.8, DEC=+60:07:14.5, via an outlier-selection procedure applied to the SDSS-DR12 photometric footprint. Similar to other recent, wide-field photometric searches, this search did not rely on previous spectroscopic or UV excess information. This approach enables the discovery of systems with sources at higher redshifts than typically probed, and with appreciable differential reddening by the lens galaxy, similar to the recently discovered DES~J0408-5354 \citep{lin17,agn17}.

Discovery NOT-ALFOSC data confirmed this system as a lens with $z_{s}=2.74$ and four images on a fold configuration. The lens redshift is $z_{l}=0.407$ from follow-up Keck-ESI spectroscopy, which together with the lens model results with $\theta_{\rm E}=(1.80\pm0.10)^{\ase}$ predicts a velocity dispersion $\sigma=(290\pm9)$km/s. Saddle-point images C,D are significantly reddened by the lens galaxy, while the A/B flux ratios agree with predictions by the lens model within $\approx0.1$mag. Microlensing is evident in the differential magnification of lines and continua. The expected time-delays (tab.~1) can be measured accurately with high-cadence campaigns spanning one or two monitoring seasons. With current data, significant monopole-quadrupole degeneracies arise in the lens model, but they can be broken with deeper and higher-resolution follow-up.

\section*{Acknowledgments}

C.G. and M.B. acknowledge support by VILLUM FONDEN Young Investigator Programme through grant no. 10123.
TJ acknowledges support provided by NASA through Program \# HST-HF2-51359 through a grant from the Space Telescope Science Institute, which is operated by the Association of Universities for Research in Astronomy, Inc., under NASA contract NAS 5-26555. TT acknowledges support by the Packard Foundation through a Packard Research Fellowship and by the National Science Foundation through grant AST-1450141. S.H.S. gratefully acknowledges support from the Max Planck Society through the Max Planck Research Group.

The data presented here were obtained in part with ALFOSC, which is provided by the Instituto de Astrofisica de Andalucia (IAA) under a joint agreement with the University of Copenhagen and NOTSA. We thank R.T.~Rasmussen and T.~Pursimo for support at the NOT, and H.~Fischer for being the lucky charm.

The ESI data presented herein were obtained at the W.M. Keck Observatory, which is operated as a scientific partnership among the California Institute of Technology, the University of California and the National Aeronautics and Space Administration. The Observatory was made possible by the generous financial support of the W.M. Keck Foundation. The authors wish to recognize and acknowledge the very significant cultural role and reverence that the summit of Mauna Kea has always had within the indigenous Hawaiian community.  We are most fortunate to have the opportunity to conduct observations from this mountain, and we respectfully say mahalo. 


\label{lastpage}


\begin{thebibliography}{}

\bibitem[Abazajian et al.(2009)]{aba09} Abazajian, K.~N., Adelman-McCarthy, J.~K., Ag\"{u}eros, M.~A., et al.\ 2009, \apjs, 182, 543
\bibitem[Agnello et al.(2013)]{agn13} Agnello, A., Auger, M.~W., \& Evans, N.~W.\ 2013, \mnras, 429, L35 
\bibitem[Agnello et al.(2015a)]{agn15} Agnello, A., Kelly, B.~C., Treu, T., \& Marshall, P.~J.\ 2015, \mnras, 448, 1446 
\bibitem[Agnello et al.(2016)]{agn16} Agnello, A., Sonnenfeld, A., Suyu, S.~H., et al.\ 2016, \mnras, 458, 3830 
%
\bibitem[Agnello et al.(2017)]{agn17} Agnello, A., Lin, H., Buckley-Geer, L., et al.\ 2017, MNRAS sumb. arXiv:1702.00406
%
\bibitem[Assef et al.(2013)]{ass13} Assef, R.~J., Stern, D., Kochanek, C.~S., et al.\ 2013, \apj, 772, 26 
%
\bibitem[Barnab{\`e} et al.(2011)]{bar11} Barnab{\`e}, M., Czoske, O., Koopmans, L.~V.~E., Treu, T., \& Bolton, A.~S.\ 2011, \mnras, 415, 2215 
\bibitem[Braibant et al.(2016)]{bra16} Braibant, L., Hutsem{\'e}kers, D., Sluse, D., \& Anguita, T.\ 2016, \aap, 592, A23 
\bibitem[Chan et al.(2015)]{cha15} Chan, J.~H.~H., Suyu, S.~H., Chiueh, T., et al.\ 2015, \apj, 807, 138 
\bibitem[Dalal \& Kochanek\ (2002)]{dal02} Dalal, N., \& Kochanek, C.~S.\ 2002, \apj, 572, 25 
\bibitem[Ding et al.(2017)]{din17} Ding, X., Liao, K., Treu, T., et al.\ 2017, \mnras, 465, 4634 
\bibitem[Falco et al.(1999)]{fal99} Falco, E.~E., Impey, C.~D., Kochanek, C.~S., et al.\ 1999, \apj, 523, 617 
\bibitem[Grillo et al.(2008)]{gri08} Grillo, C., Lombardi, M., \& Bertin, G.\ 2008, \aap, 477, 397 
\bibitem[Grillo et al.(2015)]{gri15} Grillo, C., Suyu, S.~H., Rosati, P., et al.\ 2015, \apj, 800, 38 
%
\bibitem[Inada et al.(2012)]{ina12} Inada, N., Oguri, M., Shin, M.-S., et al.\ 2012, \aj, 143, 119 
\bibitem[Jee et al.(2016)]{jee16} Jee, I., Komatsu, E., Suyu, S.~H., \& Huterer, D.\ 2016, \jcap, 4, 031 
\bibitem[Kassiola \& Kovner(1993)]{kas93} Kassiola, A., \& Kovner, I.\ 1993, Liege International Astrophysical Colloquia, 31, 571 
\bibitem[Kochanek(2006)]{koc06} Kochanek, C.~S.\ 2006, Saas-Fee Advanced Course 33: Gravitational Lensing: Strong, Weak and Micro, 91 
\bibitem[Lin et al.(2017)]{lin17} Lin, H., Buckley-Geer, E., Agnello, A., et al.\ 2017, ApJL subm., arXiv:1702.00072 
\bibitem[Mediavilla et al.(2005)]{med05} Mediavilla, E., Mu{\~n}oz, J.~A., Kochanek, C.~S., et al.\ 2005, \apj, 619, 749 
\bibitem[More et al.(2016)]{mor16} More, A., Lee, C.-H., Oguri, M., et al.\ 2016, arXiv:1608.06288 
\bibitem[Nierenberg et al.(2014)]{nie14} Nierenberg, A.~M., 
Treu, T., Wright, S.~A., Fassnacht, C.~D., \& Auger, M.~W.\ 2014, \mnras, 442, 2434 
%
\bibitem[Oguri et al.(2006)]{ogu06} Oguri, M., Inada, N.,  Pindor, B., et al.\ 2006, \aj, 132, 999 
\bibitem[Oguri \& Marshall (2010)]{om10} Oguri, M., \& Marshall, P.~J.\ 2010, \mnras, 405, 2579 
\bibitem[Ostrovski et al.(2017)]{ost17} Ostrovski, F., McMahon, R.~G., Connolly, A.~J., et al.\ 2017, \mnras, 465, 4325 
\bibitem[Paraficz \& Hjorth(2009)]{par09} Paraficz, D., \& Hjorth, J.\ 2009, \aap, 507, L49 
\bibitem[Refsdal(1964)]{ref64} Refsdal, S.\ 1964, \mnras, 128, 307 
\bibitem[Rusu et al.(2014)]{rus14} Rusu, C.~E., Oguri, M., Minowa, Y., et al.\ 2014, \mnras, 444, 2561 
\bibitem[Schechter et al.(2014)]{sch14} Schechter, P.~L., Pooley, D., Blackburne, J.~A., \& Wambsganss, J.\ 2014, \apj, 793, 96 
\bibitem[Schechter et al.(2016)]{sch16} Schechter, P.~L., Morgan, N.~D., Chehade, B., et al.\ 2016, arXiv:1607.07476 
\bibitem[Sonnenfeld et al.(2015)]{son15} Sonnenfeld, A., Treu, T., Marshall, P.~J., et al.\ 2015, \apj, 800, 94 
\bibitem[Suyu \& Halkola(2010)]{suy10} Suyu, S.~H., \& Halkola, A.\ 2010, \aap, 524, A94 
\bibitem[Suyu et al.(2012)]{suy12} Suyu, S.~H., Hensel, S.~W., McKean, J.~P., et al.\ 2012, \apj, 750, 10  
\bibitem[Suyu et al.(2014)]{suy14} Suyu, S.~H., Treu, T., Hilbert, S., et al.\ 2014, \apjl, 788, L35  
\bibitem[Treu \& Koopmans(2004)]{tk04} Treu, T., \& Koopmans, L.~V.~E.\ 2004, \apj, 611, 739
\bibitem[Treu et al.(2005)]{tre05} Treu, T., Ellis, R.~S., Liao, T.~X., et al.\ 2005, \apj, 633, 174 
\bibitem[Williams et al.(2017)]{wil17} Williams, P., Agnello, A., \& Treu, T.\ 2017, \mnras, 466, 3088 
\bibitem[Wright et al.(2010)]{wri10} Wright, E.~L., Eisenhardt, P.~R.~M., Mainzer, A.~K., et al.\ 2010, \aj, 140, 1868-1881 



\end{thebibliography}
\end{document}